\documentclass[letterpaper]{article} 
\usepackage{CameraReady/LaTeX/aaai2026}  
\usepackage{amsmath}
\usepackage{array}
\usepackage{booktabs}
\usepackage{multirow}
\usepackage{times}  
\usepackage{helvet}  
\usepackage{courier}  
\usepackage[hyphens]{url}  
\usepackage{graphicx} 
\usepackage{natbib}  
\usepackage{caption} 
\usepackage{algorithm}
\usepackage{algorithmic}

\usepackage{newfloat}
\usepackage{listings}
\DeclareCaptionStyle{ruled}{labelfont=normalfont,labelsep=colon,strut=off} 
\floatstyle{ruled}
\newfloat{listing}{tb}{lst}{}
\floatname{listing}{Listing}
\title{KGV: Integrating Large Language Models with Knowledge Graphs for Cyber Threat Intelligence Credibility Assessment}
\author{
    Zongzong Wu\textsuperscript{\rm 1},
    Fengxiao Tang\textsuperscript{\rm 1},
    Ming Zhao\textsuperscript{\rm 1},
    Yufeng Li\textsuperscript{\rm 1}
}
\affiliations{
    \textsuperscript{\rm 1}School of Computer,Science and Engineering,Central South University, Changsha, China\\
    wzy\_cookie@163.com, tangfengxiao@csu.edu.cn, meanzhao@csu.edu.cn, liyufeng1027@163.com
}

\usepackage{bibentry}

\begin{document}

\maketitle

\begin{abstract}
 Cyber threat intelligence (CTI) is a crucial tool to prevent sophisticated, organized, and weaponized cyber attacks. However, few studies have focused on the credibility assessment of CTI, and this work still requires manual analysis by cybersecurity experts. In this paper, we propose Knowledge Graph-based Verifier (KGV), the first framework integrating large language models (LLMs) with simple structured knowledge graphs (KGs) for automated CTI credibility assessment. Unlike entity-centric KGs, KGV constructs paragraph-level semantic graphs where nodes represent text segments connected through similarity analysis, which effectively enhances the semantic understanding ability of the model, reduces KG density and greatly improves response speed. Experimental results demonstrate that our KGV outperforms state-of-the-art fact reasoning methods on the CTI-200 dataset, achieving a 5.7\% improvement in F1. Additionally, it shows strong scalability on factual QA and fake news detection datasets. Compared to entity-based knowledge graphs (KGs) for equivalent-length texts, our structurally simple KG reduces node quantities by nearly two-thirds while boosting precision by 1.7\% and cutting response time by 46.7\%. In addition, we have created and publicly released the first CTI credibility assessment dataset, CTI-200. Distinct from CTI identification datasets, CTI-200 refines CTI summaries and key sentences to focus specifically on credibility assessment.
\end{abstract}


\section{Introduction}

Cyber Threat Intelligence (CTI) \cite{lee2023cyber} includes the attack tactics, techniques and procedures (TTPs) used by threat actors, as well as the report writer’s analysis of the attack.
CTI helps cybersecurity organizations quickly identify cyber threats and prevent or mitigate their impact before an attack occurs.
Low-quality CTI may lead experts to misjudge, thus threatening network security. 
 These low-quality CTIs include content misreporting \cite{li2022attackg}, outdated information \cite{griffioen2020quality}, tactical analysis errors \cite{lee2023cyber}, etc. CTIs from different sources may contradict each other, and some are even fabricated.

Current research predominantly focuses on CTI identification and extraction, yet lacks credibility assessment frameworks based on heterogeneous sources. While existing CTI databases (e.g., TTPDrill \cite{husari2017ttpdrill} used a semi-structured CTI dataset and manually extracted threat actions; CyberWire \footnote{https://thecyberwire.com/} has published a large number of annotated cybersecurity news articles, and some methods \cite{satyapanich2020casie} used these articles for CTI identification.) support CTI identification tasks, their objectives diverge fundamentally from credibility assessment: the former prioritizes claim validation over adversarial behavior detection, rendering them inadequate for credibility evaluation. To address this gap, we propose CTI-200, the first heterogeneous-source dataset designed for CTI credibility assessment, supporting verification through three dimensions: attack sources, content features, and temporal attributes.

\begin{figure}[t!]
\centering
\includegraphics[width=\columnwidth]{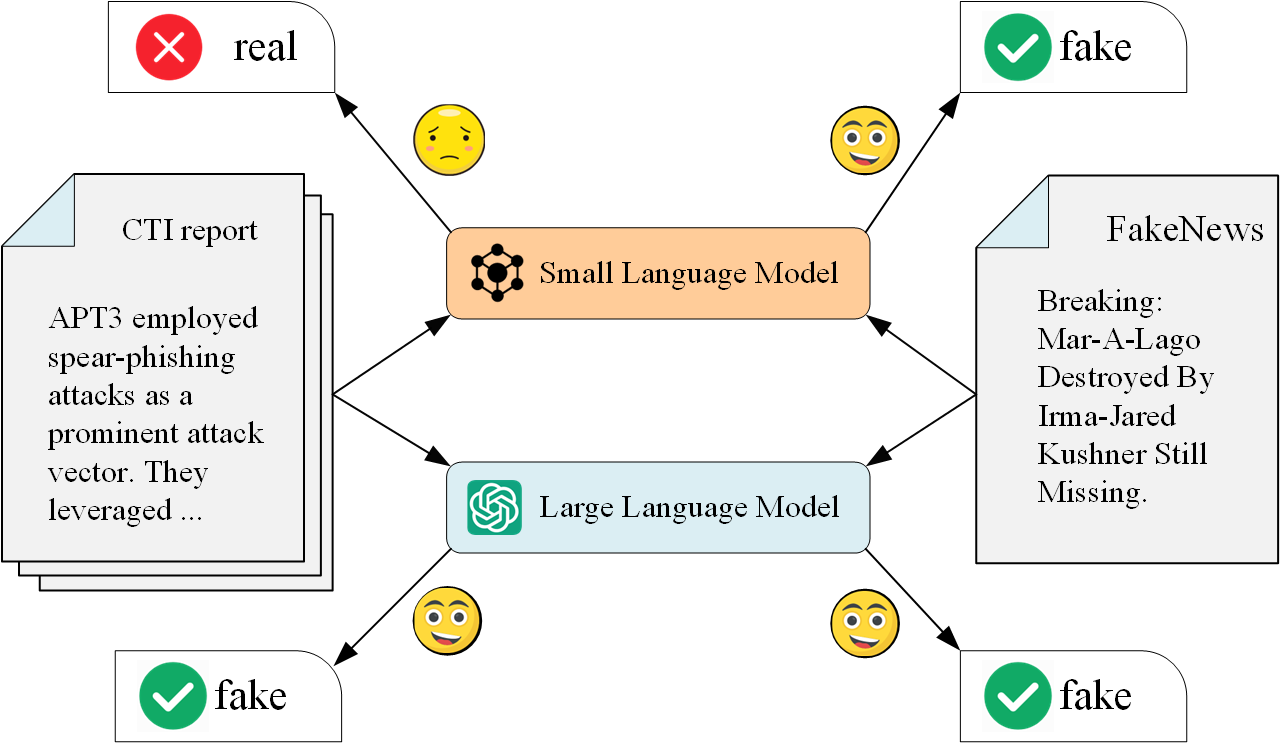}
\caption{The capabilities of large and small language models in handling different categories of tasks.}
\label{fig:introduce}
\end{figure}

Another challenge of CTI credibility assessment lies in the uniqueness of CTI content. 
  We have observed that some CTI identification methods \cite{milajerdi2019poirot} also superficially involve evaluation work, but these methods usually rely on a large number of predefined rules to evaluate a single CTI. 
For example, Gao et al. \cite{gao2021enabling} used graph mining techniques to infer the relationships between cyber threat infrastructures in a single report based on the constructed intelligence graph. 
  However, the cyber attacks described in CTI are highly characteristic, and some APT organizations have multiple code names, so these methods require pre-defining a large number of rules when building intelligence graphs, which is a labor-intensive task. 
Lin et al. \cite{lin2023correlation} evaluated CTI only on a single dimension of malicious traffic. 
  This work also clustered similar flows for each label by predefining a large number of feature rules, which also consumed a huge amount of manpower. 
In addition, the credibility assessment of network threats needs to consider multiple dimensions (such as attack source, intelligence content and release timeliness, etc.). 
  Network traffic is only a small part of intelligence content, and it is insufficient to evaluate CTI credibility only through network traffic.
   Although these studies involve the credibility assessment of CTI, they have not explored the challenges faced by this work in depth, nor have they summarized a feasible paradigm.


The primary objective of CTI credibility assessment is to verify the truthfulness of key viewpoints in CTI, which shares similarities with fake news detection. Knowledge Graph (KG)-based methods \cite{shahi2023fakekg} in fake news detection typically reason using evidence graphs with fine-grained node levels \cite{zhong2019reasoning,jin2022towards} . However, CTI is more complex, often involving longer texts and multiple claims compared to the single-claim, shorter fake news. This complexity requires stronger reasoning and richer knowledge, challenging traditional small language models that need complex labeled KGs, a labor-intensive process. Recent advancements in Large Language Models (LLMs) have enabled more efficient CTI processing. For instance, aCTIon \cite{siracusano2023time} uses GPT-3.5 to extract structured text, while Llm-Tikg \cite{hu2024llm} builds KGs on CTI  with LLMs. These works have inspired us to use LLMs for CTI credibility assessment. LLMs, due to their strong prior knowledge and text awareness, can be guided by relatively simple KGs to perform effective claim verification, mitigating issues like factual illusion by leveraging high-quality factual data from KGs. Figure \ref{fig:introduce} shows the capabilities of large and small language models in handling different categories of tasks.


In this paper, we propose Knowledge Graph-based Verifier (KGV), a novel CTI credibility assessment framework that combines KG and LLMs. 
  We introduced a set of specially designed prompts to enable LLM to automatically extract key points from CTI and decompose these key points into claims to be verified. 
The idea behind the KGV framework is to discover factual knowledge triples by performing graph retrieval on the extracted claims through LLM, and to fact-check the claims decomposed from key points by learning claim triples-factual knowledge triples pairs through LLM. 
  Our approach shows a significant difference from previous methods: by leveraging pre-trained LLM, we only need to build a factual KG with a relatively simple structure to complete our work, and effectively avoid the fact hallucination of LLMs and reduce the response time of the model. 
Specifically, our KG uses paragraphs as nodes and connects nodes by semantic similarity between paragraphs. 
  The benefits of this are: (1) it avoids complex relationship modeling and reduces computational complexity; 
(2) paragraphs usually contain complete semantic units, which can better capture the semantic relationship between paragraphs, thereby improving the contextual reasoning ability of the model. 
  In addition, we created and published the first dataset applied in the field of CTI credibility assessment. 
This dataset includes 1,000 CTIs from heterogeneous intelligence sources. 
  We divided 200 CTIs to be verified and 800 fully verified reliable CTIs, and divided these 1,000 CTIs into 200 groups, each of which contains one CTI to be verified and four related reliable CTIs as verification clues. We provide a detailed description of the dataset creation process in the appendix.

The main contributions of this paper are as follows:

1. We propose a new framework called KGV, which provides a feasible paradigm for CTI credibility assessment. 
  As far as we know, this is the first work to introduce LLMs into CTI credibility assessment. 
It is also the first work to evaluate CTI credibility from multiple dimensions.

2. We propose the integration of a KG with paragraphs as nodes into KGV, which reduces the graph's scale while providing richer semantic information. KGV can robustly extract factual knowledge to alleviate the factual hallucinations of LLMs and reduce the response time of the model.

3. The KGV framework can not only complete the CTI credibility assessment work well, but also demonstrates strong generalization ability in the field of factual QA and fake news detection.

4. We created and released the first heterogeneous source dataset for CTI credibility assessment, which contains 1,000 CTIs from heterogeneous sources and behavioral knowledge of 45 APT organizations. 
  We also evaluated the performance of our KGV framework on this dataset.

\section{Related work}

\subsection{CTI credibility assessment} 

Current research on the multi-dimensional evaluation of heterogeneous source data remains in its nascent stages, with limited studies investigating the assessment of network threat intelligence (CTI) from multiple dimensions. The primary focus of researchers has been on the identification and classification of threat intelligence. A sparse number of studies have delved into evaluating the authenticity of threat intelligence from specific dimensions or assessing its reliability by identifying false positives within individual intelligence reports. For instance, AttacKG \cite{li2022attackg} automatically extracts structured attack behavior graphs from CTI reports and identifies relevant attack techniques.
MultiKG \cite{wang2024multikg} leverages LLMs to automate the analysis, construction, and merging of attack graphs from multiple sources, generating a fine-grained, multi-source attack knowledge graph.


These rudimentary approaches do not fully encapsulate the evolving trajectory of CTI trustworthiness assessment. Our observations indicate a significant lack of standardized guidelines in current research. 
In response, we propose a holistic evaluation methodology that leverages multiple CTI reports from heterogeneous sources to assess individual CTI reports. 

\subsection{Factual Reasoning}

\textbf{Fact checking} commonly used scenarios for fake news detection, these works infer the authenticity of news descriptions based on single-modal or multi-modal data such as text content\cite{qi2019exploiting} and images\cite{qi2021improving}.Some studies\cite{tseng2022kahan,hu2021compare} have also introduced external knowledge bases based on KGs to supplement news content. These studies all use pre-trained small language models for text reasoning, but few consider the potential of LLMs for factual reasoning.

\subsection{CTI dataset} 
To the best of our knowledge, no datasets specifically designed for the CTI credibility assessment have been developed prior to this study. Existing CTI datasets are primarily intended for tasks such as the identification and extraction of threat behaviors. For instance, TTPDrill \cite{husari2017ttpdrill} utilizes a semi-structured Symantec dataset to extract threat actions; Extractor \cite{satvat2021extractor} accounts for heterogeneous data sources, and Open-CyKG \cite{sarhan2021open} employs a CTI dataset focused on malware. However, these datasets are solely created for the purpose of identifying malicious activities, with label types limited to simple annotations such as entity types and entity relationships, rendering them unsuitable for CTI trustworthiness evaluation. The absence of appropriate datasets for training evaluation models is a significant reason why CTI credibility assessments continue to rely heavily on manual expert analysis.


\section{Method}
\begin{figure*}[ht]
 \includegraphics[width=\textwidth]{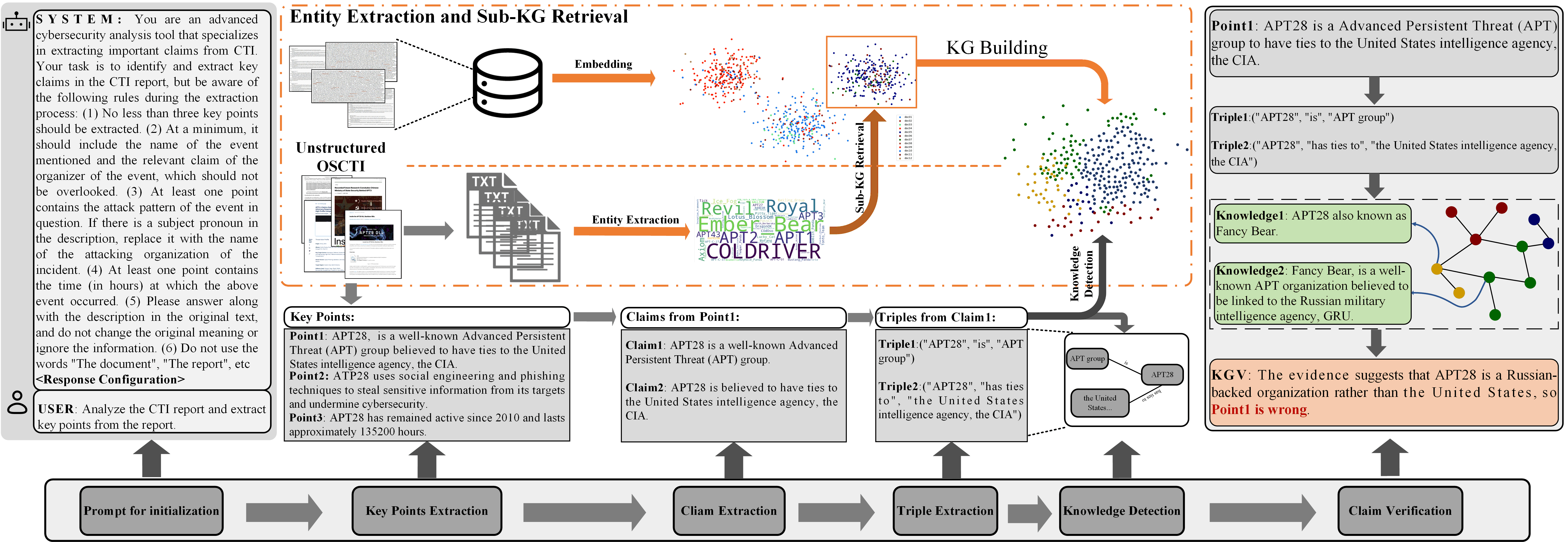}
  \caption{Overview of KGV. The KGV framework mainly includes two
cold start stages: KG Building, entity extraction and sub-KG retrieval, and six main stages: prompts for initialization , key points
extraction, claim extraction, triple extraction, knowledge detection,
and claim verification. }
  \label{overviewKGV}
\end{figure*}

As shown in Figure \ref{overviewKGV}, faced with a CTI to be verified, our framework enables LLM to extract key viewpoints of concern to cybersecurity experts through specially designed prompts, including: the threat actor (attack source), the method used by the threat attack, and the timeliness of the threat attack. 

  Next, the LLMs will extract statements from the key viewpoints, and our work mainly focuses on fact-checking multiple statements extracted from the reports to be verified. 
It is worth noting that we do not evaluate the overall quality of CTI, because the overall quality of the text is highly subjective and such an evaluation would be unfair. 
  We only conduct authenticity assessments on the claims proposed in the CTI to be verified, and use the results to guide relevant practitioners to understand the overall credibility of the CTI.
  Next, we will introduce each stage in detail. 


\subsection{KG Building}

We first construct a KG based on multiple clue CTIs in the CTI-200 dataset, and use it as an external knowledge base to guide the LLMs to evaluate the credibility of CTI.  Specifically:

Assume the total number of clue CTIs: \( \mathrm{R}=\left\{R_{1}, R_{2}, R_{3}, \ldots, R_{N}\right\} \), then the number of paragraphs contained in each clue CTI \(R_i\) is \( P_{R_i} = \left\{ P_{1}, P_{2}, P_{3}, \ldots, P_{M} \right\} \). From this we can know the total number of paragraphs \(S\):
Q
\begin{equation}
    S={\textstyle \sum_{i=1}^{N}}{\textstyle  \ P_{R_i} }
\end{equation}

Then the number of nodes in the KG is $V= \left\{ v_i \mid i = 1, 2, \dots, S \right\}$.
In order to reflect the semantic similarity between related paragraphs, we establish the relationship between paragraphs through CTI paragraph segmentation and paragraph-paragraph semantic relationship modeling. 
  Specifically, when chunking CTI, we record the order of paragraphs and connect the physically adjacent paragraphs in the document. 
In addition, because multiple CTIs may contain records of attacks on the same organization, we extract keywords from paragraphs by prompting LLM, connect paragraphs containing the same keywords, and calculate the semantic similarity between paragraphs for connection. 
  Therefore, related content blocks are connected, which can better facilitate graph retrieval.

Specifically, we choose Bert \cite{devlin2018bert} as the embedding model and perform text preprocessing on each paragraph, including word segmentation, stop word removal, and stem extraction to reduce noise interference. 
  Each paragraph is converted into a high-dimensional vector representation by taking the average of the vectors of all tokens in the paragraph:
\begin{equation}
    \varphi (p) = \frac{1}{|p|} {\textstyle \sum_{i=1}^{|p|}}BERT(p,i)  
\end{equation}

Among them, \(BERT(p,i)\) is the embedding of the \(i-th\) token in paragraph \(p\), and \(|p|\) is the number of tokens in paragraph \(p\). This vector can capture the semantic dependency between paragraph texts. The semantic similarity between paragraphs is calculated by the following formula:

\begin{equation}
\operatorname{Cosine} \operatorname{Similarity}\left(\varphi\left(p_{i}\right), \varphi\left(p_{j}\right)\right)=\frac{\varphi\left(p_{i}\right) \cdot \varphi\left(p_{j}\right)}{\left\|\varphi\left(p_{i}\right)\right\|\left\|\varphi\left(p_{j}\right)\right\|}
\end{equation}

Among them, \(\left\|\varphi\left(p_{i}\right)\right\|\) and \(\left\|\varphi\left(p_{j}\right)\right\|\) represent the norm (modulus) of the embedding vector \(\varphi\left(p_{i}\right)\) of paragraph \(i\) and the embedding vector \(\varphi\left(p_{j}\right)\) of paragraph \(j\), respectively. This is done by comparing the embedding vectors of two paragraphs using cosine similarity, where values closer to 1 indicate that the two paragraphs are more similar.In this paper, we choose to concatenate two paragraphs if their cosine similarity is greater than 0.8, and explore this choice in ablation study.

\subsection{Entity Extraction and Sub-KG Retrieval}
The first is text entity extraction, which aims to enable LLMs to quickly link the CTI  to be detected with the factual knowledge in the knowledge graph to facilitate the cold start of LLMs on the KG. 
  Based on this, we should pay attention to two points: (1) retrieve the factual knowledge about CTI  in the knowledge graph as comprehensively as possible; (2) At the same time, it is necessary to avoid retrieving some noise knowledge that is irrelevant to the description in CTI  as much as possible. 
Therefore, the selection of entities should be highly directional, because there are many entities in CTI . If all entities are fed into the sub-KG retrieval without screening, LLMs are likely to produce fact hallucinations and easily retrieve noisy knowledge. 
  In this paper, we uniformly use attack organizations as contact entities. This is determined by the characteristics of CTI reports. 
    We observe that almost all CTI reports mention the corresponding attack organizations when describing attack incidents. 
By selecting the attack organization as the entity to be extracted, we can ensure that as many relevant local sub-KGs in the KG as possible are called, which means more useful factual knowledge triples, while avoiding the intrusion of irrelevant noise knowledge as much as possible. 
  

\subsection{Prompts for initialization and key point extraction}
When faced with a CTI report to be verified, we generate corresponding key viewpoints from the LLMs. It is worth noting that CTI expresses many opinions, but not all of them are relevant to the quality assessment of CTI. This work mainly evaluates the quality of CTI from three dimensions: attack source, attack method and attack timeliness.Therefore, we use the following instructions to make the LLMs generate specific declarations:

\textbf{LLMs prompts: You are an advanced cybersecurity analysis tool that specializes in extracting important claims from
CTI. Your task is to identify and extract key claims in the
CTI report, but be aware of the following rules during the
extraction process: (1) No less than three key points should
be extracted. (2) At a minimum, it should include the name of
the event mentioned and the relevant claim of the organizer
of the event, which should not be overlooked. (3) At least one
point contains the attack pattern of the event in question. If
there is a subject pronoun in the description, replace it with
the name of the attacking organization of the incident. (4)
At least one point contains the time (in hours) at which the
above event occurred. (5) Please answer along with the de-
scription in the original text, and do not change the original
meaning or ignore the information. (6) Do not use the words
``The document'', ``The report'', etc. }


\subsection{Claim Extraction}
After identifying the key points of CTI, the LLMs will extract all claims related to these points. The underlying rationale is that these key points extracted by LLMs often contain multiple factual claims that require verification. For example, 
the core idea to be verified is: ``APT28 uses social engineering techniques to steal sensitive information from its targets.'' Two claims within this idea require verification: ``APT28 uses social engineering techniques to steal sensitive information from its targets'' and ``APT28 uses phishing techniques to steal sensitive information from its targets.''

With only partial factual knowledge triples—(``phishing techniques,'' ``aimed at,'' ``stealing sensitive information'')—as support, if LLMs attempt to verify the entire idea directly, they are likely to misinterpret the reasoning process and incorrectly assume that the entire idea is supported. However, these factual knowledge triples only support the claim ``APT28 uses phishing techniques to steal sensitive information from its targets'' and do not support the claim ``APT28 uses social engineering techniques to steal sensitive information from its targets.'' To enable LLMs to verify these claims separately, we decompose the core ideas into more fine-grained claims. 


\subsection{Triple extraction and knowledge detection}
After obtaining the claim, we guide LLMs to extract entity tuples from the claim. Some methods \cite{sancheti2024llm} propose to extract entities directly from the claim to match the factual knowledge triples. However, only through entities, many triples will be matched from the factual knowledge. More triples mean that LLMs will have a larger fact hallucination space. LLMs have powerful semantic understanding capabilities and support triple-triplet matching to obtain more accurate factual knowledge as guidance. 


\subsection{Point verification}
Given the acquired claim triples, we use LLM to compare the proposed claim with the factual information in KG, so that the authenticity of the statement is given by LLM, and the authenticity of the point is verified by combining multiple claims contained in a point. 

  The main idea behind this is that LLM fully verifies the facts of each claim. 
Only local verification may not convince LLM to thoroughly verify the authenticity of the point, thus creating a factual illusion. 

\section{Experiments}
We evaluate the KGV framework and other natural language reasoning models at different levels on the CTI-200 dataset. These models include factual reasoning models based on small language models, such as HAN \cite{ma2019sentence}, EHIAN \cite{wu2021evidence}, MAC \cite{vo2021hierarchical}, GET \cite{xu2022evidence}, MUSER \cite{liao2023muser}, ProgramFC\cite{pan2023fact},GAMC\cite{yin2023gamc}, FinerFact\cite{liu2023finerfact} and SheepDog\cite{wu2023sheepdog}, as well as factual reasoning models based on large language models, including GPT-3.5\cite{ouyang2022training},LLaMa 3.1 \cite{dubey2024llama},GPT-4 \cite{achiam2023gpt}, DPR \cite{karpukhin2020dense}, IRCoT \cite{trivedi2022interleaving}, CoT \cite{wei2022chain},QKR \cite{baek2023knowledge}, KGR \cite{guan2024mitigating}, and KGP-T5 \cite{wang2024knowledge}. 

\subsection{Database and Experimental settings}
\textbf{CTI-200} is a small dataset we created, focusing on CTI credibility evaluation. We describe the creation process of this dataset in detail in the Appendix. We divided the 200 CTIs in the CTI-200 dataset into CTIs that need verification and factual CTIs. For the CTIs requiring verification, we used a large language model with specific prompts to extract opinions that need to be verified as questions, resulting in a total of 864 questions. For factual CTIs, we segmented the data and constructed a factual KG as described in Appendix.



\textbf{Politifact} is a dataset consisting of 19,341 pairs of original tweets and fact-checked articles from snopes.com and politifact.com. Each pair is labeled as 1 if the article fact-checked the tweet. After labeling, the final dataset contains 13,239 positive pairs.

\textbf{GossipCop} is a dataset containing entertainment news with social context and provides fact-checked articles from the eonline website. It contains both fake news and true news, with fake news mainly related to celebrity rumors. GossipCop assigns a truthfulness rating from 0 to 10, where 0 means completely false and 10 means completely true.

\textbf{Experimental settings}
The experimental environment is based on the standard LLM inference setup, where GPT series LLMs, including GPT-3.5 and GPT-4 from the OpenAI API, and LLaMA-3.1 locally deployed with Ollama, are called through the OpenAI platform. We do not involve model training but focus on zero-shot or few-shot prompting integrated with knowledge graphs (KGs). All experiments are run on a Linux server equipped with 6 NVIDIA A100 GPUs.

\textbf{Implementation Details}
The CTI credibility assessment process is divided into a cold start phase (entity extraction and sub-KG retrieval) and a main phase (prompt initialization, key point extraction, knowledge detection, and claim verification). KG construction uses similarity analysis, with the similarity threshold set to 0.8 to ensure semantic relevance of the nodes. We use the iterative process in Algorithm 1 for validation scoring. Full process examples can be found in Appendix Tables 9 and 10. The experiments are repeated 3 times, and the average value and standard deviation are reported.

\subsection{Overall Results}
\begin{figure}[htbp]
\centering
\setlength{\abovecaptionskip}{0.2cm}
\setlength{\belowcaptionskip}{-0.1cm} 
\includegraphics[width=\columnwidth]{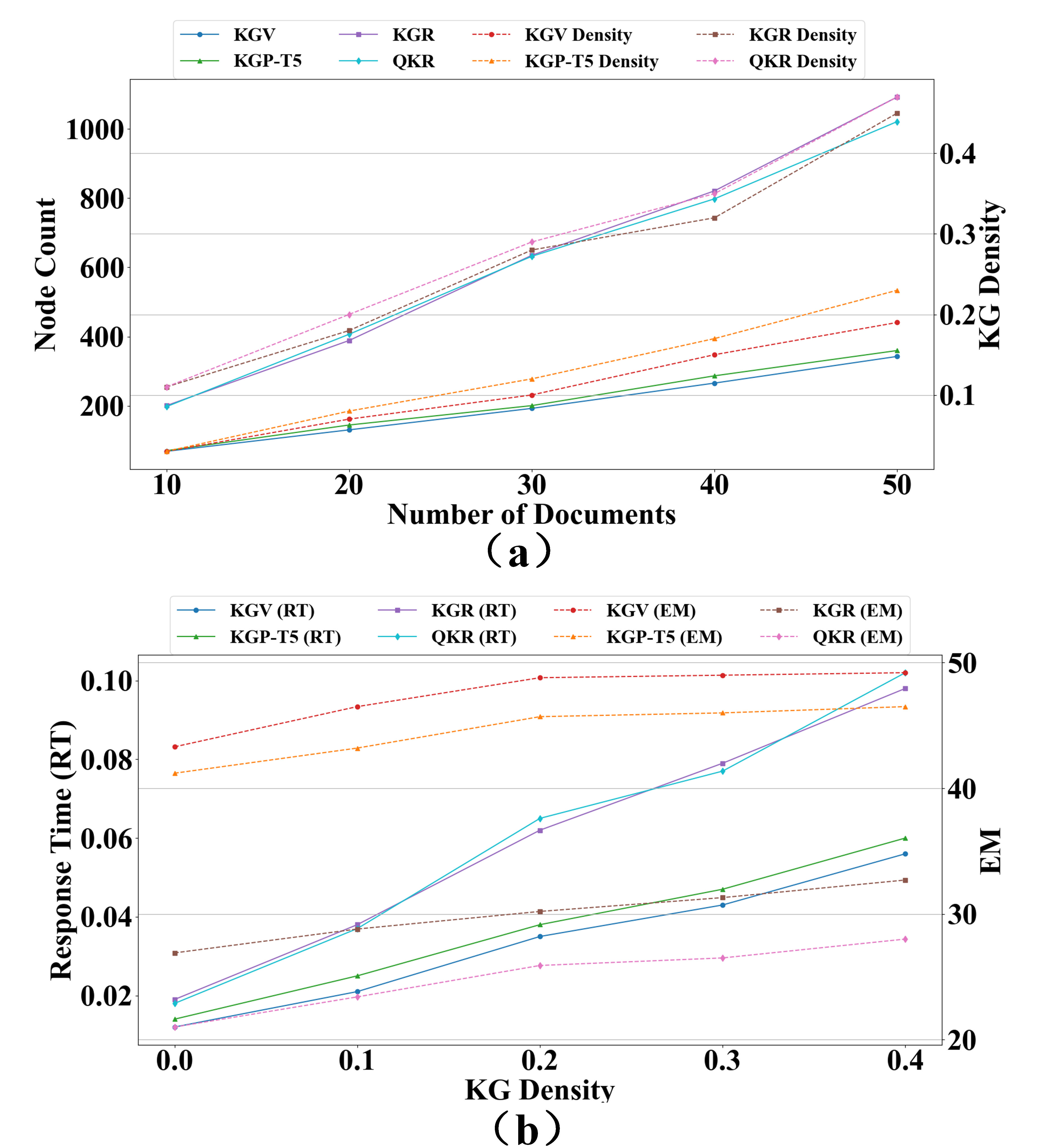}
\caption{(a) Comparison of the number of nodes and KG density; (b) Comparison of response time and EM at the same KG density.}
\label{fig:em-time-density-count}
\end{figure}

We compare our method with recently popular factual reasoning approaches. Additionally, we evaluate the generalization ability of our KGV by comparing it with current state-of-the-art factual question-answering methods based on LLMs. The experimental results are shown in Tables \ref{tab:t1}, and Figure \ref{fig:em-time-density-count}. Our method shows significant advantages in both factual reasoning and factual question answering based on LLMs.

\begin{table*}[t]
    \centering
    \caption{Performance comparison on CTI-200, PolitiFact, and GossipCop}
    \label{tab:t1}
    \begin{tabular}{>{\centering\arraybackslash}m{2cm}ccc|ccc|ccc}
        \toprule
        \multirow{2}{*}{\centering\textbf{Method}} & \multicolumn{3}{c}{\textbf{CTI-200}} & \multicolumn{3}{c}{\textbf{PolitiFact}} & \multicolumn{3}{c}{\textbf{GossipCop}} \\
        \cmidrule(lr){2-4} \cmidrule(lr){5-7} \cmidrule(lr){8-10}
        & \textbf{P} & \textbf{R} & \textbf{F1} 
        & \textbf{P} & \textbf{R} & \textbf{F1}
        & \textbf{P} & \textbf{R} & \textbf{F1} \\
        \midrule
        HAN         & 0.625 & 0.647 & 0.636 & 0.676 & 0.682 & 0.679 & 0.721 & 0.716 & 0.678 \\
        EHIAN       & 0.617 & 0.768 & 0.684 & 0.680 & 0.651 & 0.674 & 0.713 & 0.749 & 0.673 \\
        MAC         & 0.700 & 0.686 & 0.687 & 0.695 & 0.704 & 0.700 & 0.682 & 0.662 & 0.672 \\
        GET         & 0.721 & 0.694 & 0.705 & 0.712 & 0.770 & 0.725 & 0.751 & 0.749 & 0.727 \\
        MUSER       & 0.791 & 0.740 & 0.817 & 0.735 & 0.780 & 0.757 & 0.784 & 0.843 & 0.734 \\
        ProgramFC   & 0.795 & 0.725 & 0.758 & 0.805 & 0.775 & 0.790 & 0.792 & 0.763 & 0.777 \\
        GAMC        & {0.853} & {0.877} & {0.864} 
        & {0.867} & {0.798} & {0.831} & {0.936} & {0.949} & {0.943} \\
        FinerFact        & {0.845} & {0.837} & {0.917} 
        & {0.903} & {0.928} & {0.917} & {0.870} & {0.867} & {0.868} \\
        SheepDog    & {0.875} & {0.834} & {0.854} 
        & {0.902} & {0.869} & {0.884} & {0.814} & {0.707} & {0.758} \\
        \midrule
        GPT-3.5     & 0.897 & 0.875 & 0.886 & 0.857 & 0.879 & 0.878 & 0.897 & 0.875 & 0.886 \\
        LLaMa 3.1 (405B)& 0.922 & 0.888 & 0.901 & 0.911 & 0.900 & 0.906 & 0.912 & 0.872 & 0.892 \\
        GPT-4      & 0.936 & 0.913 & 0.924 & 0.920 & 0.863 & 0.890 & 0.928 & 0.905 & 0.916 \\
        \midrule
        KGV (OURS)  & \textbf{0.987} & \textbf{0.975} & \textbf{0.981} & \textbf{0.974} & \textbf{0.950} & \textbf{0.962} & \textbf{0.963} & \textbf{0.975} & \textbf{0.969} \\
        \bottomrule
    \end{tabular}
\end{table*}


 As shown in Table \ref{tab:t1}, our framework has achieved significant improvements in CTI credibility assessment compared to mainstream factual reasoning methods, demonstrating a 22.3\% improvement in F1 over the most advanced method (MUSER) based on a small language model. This result highlights the superior effectiveness of our LLMs-based KGV method in fact reasoning and verification. LLMs possess strong priors on textual knowledge and a keen perception of contextual clues, allowing external knowledge bases with simpler structures to remain reliable, unlike small language models that may introduce noise from external knowledge bases. Moreover, our KGV method enhances F1 by at least 5.7\% compared to other LLMs-based methods. LLMs are susceptible to factual hallucinations, especially in multi-hop factual reasoning tasks for complex problems; however, our KGV method mitigates this issue by integrating knowledge graphs with high-quality factual content, effectively reducing factual inaccuracies and achieving superior performance.



In general, our KGV framework demonstrates strong performance on multiple datasets. It also shows robust scalability in other factual reasoning tasks.

\subsection{Ablation study}
To validate the effectiveness of our KGV framework, we conduct an extensive exploration of various settings in KGV, aiming to answer the following research questions:

\textbf{RQ1}: Why do we propose the KG with such a simple structure introduced in the text to guide LLMs to perform factual reasoning and verification work?

\textbf{RQ2}: Why do we choose to connect paragraphs with semantic similarity greater than 0.8 in the process of constructing KG?

\textbf{RQ3}: Why do we need to decompose points into factual claims for fact checking?

\textbf{RQ4}: Why do we extract claim triples instead of directly retrieving factual knowledge through entities?

\subsubsection{KG Comparison  (\textbf{RQ1},\textbf{RQ2})}

We first discuss why KG is introduced to guide LLMs. As shown in Table \ref{tab:t3}, we compare the performance of KGV with and without KG guidance on the CTI-200 dataset. We can clearly observe that when dealing with complex multi-hop reasoning problems, the KGV framework guided by KG is more likely to get rid of the fact hallucination problem of LLM and improves F1 by 6.4\%.

\begin{table}[h]
    \centering
    \caption{with KG vs without KG}
    \label{tab:t3}
    \begin{tabular}{>{\centering\arraybackslash}m{2cm}ccc}
        \toprule
        \multirow{2}{*}[-1ex]{\centering\textbf{Method}} & \multicolumn{3}{c}{\textbf{CTI-200}} \\
        \cmidrule(lr){2-4}
        & \textbf{Precision} & \textbf{Recall} & \textbf{F1 Score} \\
        \midrule
        Without KG         & 0.912 & 0.924 & 0.918 \\
        With KG   & \textbf{0.987} & \textbf{0.975} &\textbf{0.981} \\
        \bottomrule
    \end{tabular}
\end{table}

Next, we explore whether the proposed KG with a simple structure has advantages over the traditional KG with entities as nodes. As shown in Table \ref{tab:t3}, we compare the number of nodes, verification accuracy, and response speed generated by two different KGs. Experiments show that the KG structure we proposed with paragraphs as nodes integrates more factual knowledge while generating fewer nodes, which reduces the KG density and improves the efficiency of graph traversal. Therefore, we can observe that the simple structure KG proposed by KGV not only brings performance improvement, but also greatly improves the response speed of the model.

\begin{table}[h]
    \centering
    \caption{Comparison of KG node types}
    \label{tab:t4}
    \begin{tabular}{>{\centering\arraybackslash}m{2cm} >{\centering\arraybackslash}m{1.5cm} >{\centering\arraybackslash}m{1.5cm} >{\centering\arraybackslash}m{1.5cm}} 
        \toprule
        \multirow{2}{*}[-2ex]{\vspace{1.5ex}\textbf{Node type}} & \multicolumn{3}{c}{\textbf{CTI-200}} \\
        \cmidrule(lr){2-4}
        & \textbf{Precision} & \textbf{Node count} & \textbf{Response time(s)} \\
        \midrule
        Entity & {0.958} & 1034 & 0.12 \\ 
        Paragraph & \textbf{0.987} & \textbf{364} & \textbf{0.056} \\
        \bottomrule
    \end{tabular}
\end{table}

\begin{figure}[H]
\vspace{-0.3cm}
\setlength{\abovecaptionskip}{0.2cm}
\setlength{\belowcaptionskip}{-0.1cm} 
\centering
\includegraphics[width=\columnwidth]{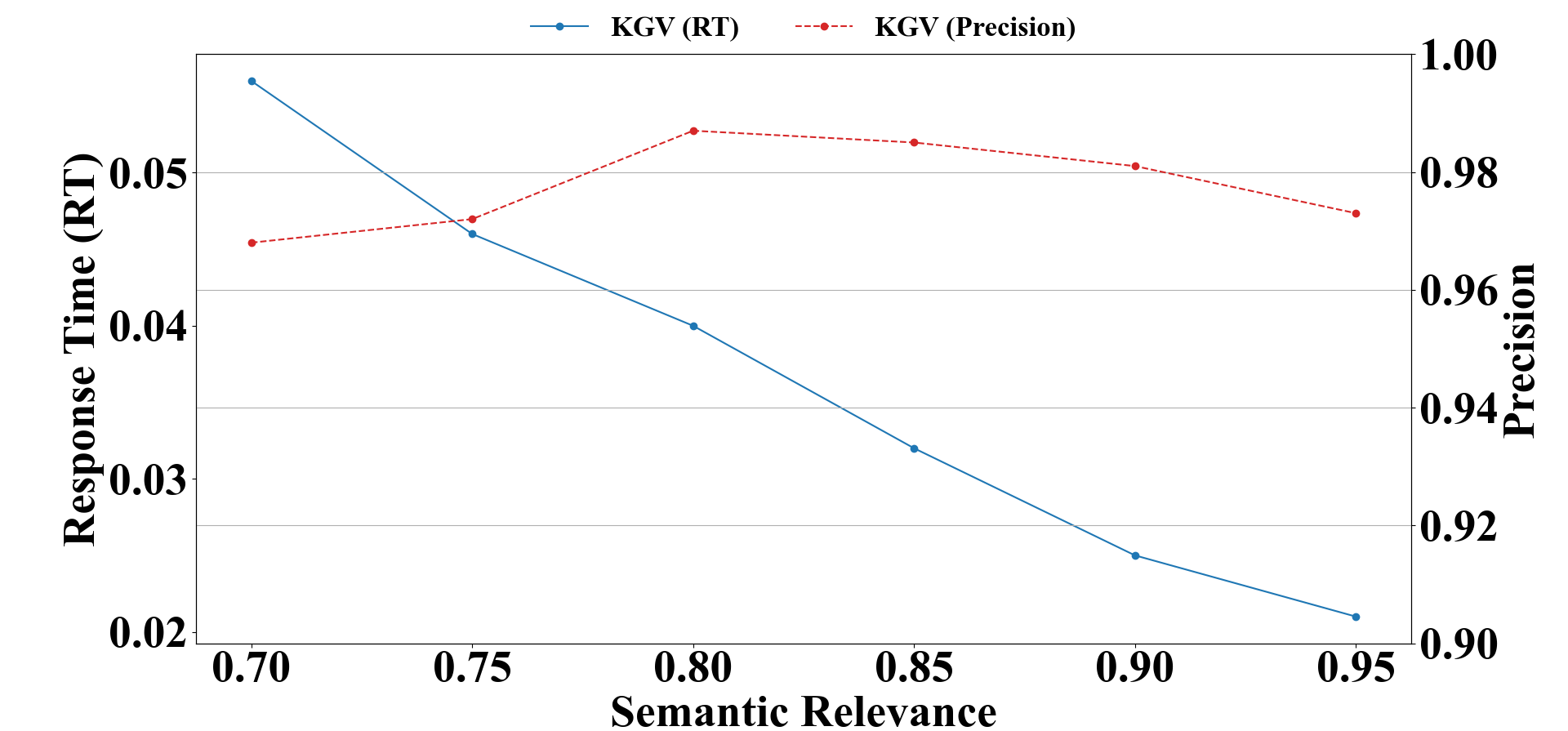}
\caption{Comparison of response time and KGV precision for semantic relevance threshold selection.}
\label{fig:Semantic Relevance}
\end{figure}

 Finally, we explored the node connection scheme during KG construction. 
  Specifically, we explore the impact of the threshold of semantic similarity between paragraphs on fact verification results on the CTI-200 dataset. 
   As shown in Figure \ref{fig:Semantic Relevance}, the most cost-effective effect is obtained when the threshold is 0.8. 
We can observe that for response time, the model becomes faster and faster in the range of 0.7-0.95. 
  This is because as the threshold increases, the density of the knowledge graph becomes smaller and smaller, so the response time becomes faster and faster. 
However, within the threshold range of 0.7-0.8, the precision of the model gradually improves. 
  This is because the lower threshold integrates more irrelevant clue knowledge into the KG, thus giving the LLM a larger fact illusion space. 
However, within the threshold range of 0.8-0.95, the precision of the model gradually decreases. 
   This is because as the threshold increases, the KG density becomes lower and lower, and LLM finds it difficult to obtain useful external knowledge when reasoning about complex multi-hop problems.

\subsubsection{KGV Process Analysis (RQ3,RQ4)}

Initially, we designed three stages for the fact verification process of KGV: key points extraction, fact knowledge retrieval, and points verification. 
  To understand the impact of each stage of the KGV method on the performance, we conducted a comprehensive analysis of the incorrectly verified cases on the CTI-200 dataset and summarized the number of incorrect cases caused by each stage in Table  \ref{tab:t5}.

\begin{table}[h]
    \centering
    \caption{The number of error cases caused by each stage of KGV-V1 and KGV-V2.}
    \label{tab:t5}
    \begin{tabular}{>{\centering\arraybackslash}m{2cm} cc}
        \toprule
        \textbf{Stage} & \textbf{KGV-V1} & \textbf{KGV-V2} \\
        \midrule
        Stage 1 & 4  & 3 \\
        Stage 2 & 72 & 7 \\
        Stage 3 & 36 & 8 \\
        Stage 4 &  / & 4 \\
        \bottomrule
    \end{tabular}
\end{table}

We found that the main reasons for verification errors are that the factual claims contained in points are not fully verified and the fact knowledge retrieval is inaccurate. 
   Subsequently, we improved the KGV design process and divided it into four stages: key point extraction, claim extraction, claim triple extraction and fact knowledge retrieval, and point verification. 
The number of error cases caused by each stage is summarized in Table \ref{tab:t5}.
(more examples are provided in Figure \ref{fig:Analysis_1}, \ref{fig:Analysis_2}, \ref{fig:Analysis_3}, \ref{fig:Analysis_4} in the Appendix)

\begin{figure}[h]
\centering
\setlength{\abovecaptionskip}{0.2cm}
\setlength{\belowcaptionskip}{-0.1cm} 
\includegraphics[width=\columnwidth]{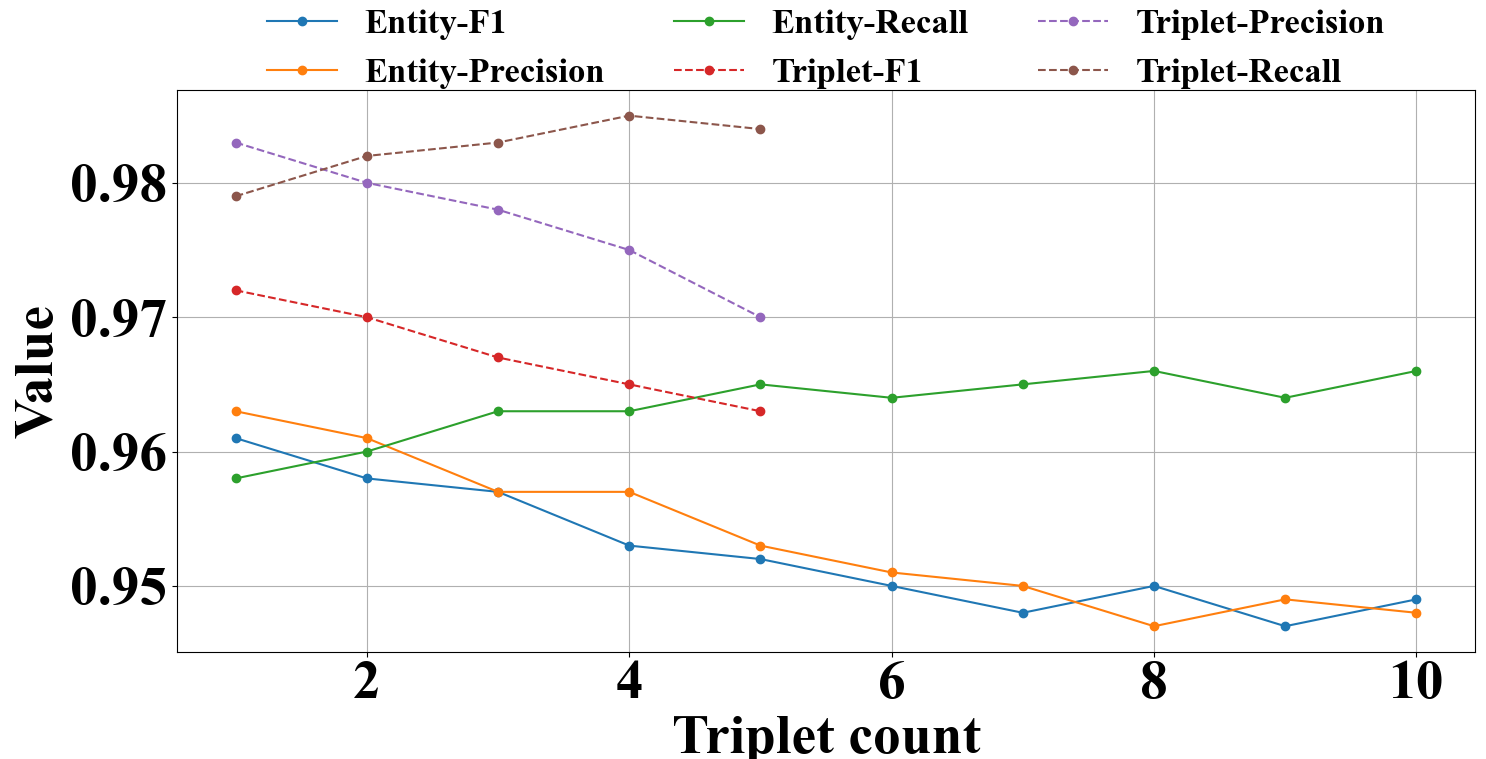} 
\caption{Impact of the number of retrieved fact triplets on verification performance.}
\label{fig:triplet entity}
\end{figure}

 In addition, we further explore the necessity of factual knowledge retrieval via claim triples. We observe that many methods including KGR propose to retrieve factual knowledge through entities in claim, but this process extracts too many useless fact triples, resulting in more noise injection. Retrieving triple-triplet pairs based on claim triples can extract more accurate and useful factual knowledge triples. As shown in Figure \ref{fig:triplet entity}, we fix the number of fact triplets retrieved by the two methods. The experimental results show that directly retrieving fact knowledge through entities shows worse results because it cannot ensure the retrieval of useful fact triplets. In addition, when the number of fact triples increases to more than 5, the number of fact triplets retrieved through triple pairs will no longer increase, but the number of fact triplets retrieved through entities will continue to rise, which will bring LMM a larger noisy fact space and lead to the problem of fact hallucination. However, the above problem can be effectively avoided by retrieving triple-triplet pairs. Finally, we present more full-process examples of KGV performing factual reasoning in Figure \ref{fig:Analysis_3} and \ref{fig:Analysis_4} of the appendix.

\section{Conclusion and Future work}
This paper proposes a KGV framework for the CTI credibility assessment task. 
  This is the first framework to deal with CTI credibility assessment, and we believe that this framework explores a feasible and efficient paradigm for CTI credibility assessment. 
Compared with other LLM factual reasoning methods, this framework effectively alleviates the factual illusion in the LLM reasoning process through a KG with a simple structure, and does not lead to insufficient integrated factual knowledge due to its simple structure. 
  Experimental results show that KGV is significantly superior to other factual reasoning methods in CTI credibility assessment, which reflects the expertise of KGV in the field of CTI credibility assessment. 

As for future work, we plan to consider more dimensions of CTI, including the ability to accurately identify and assess kill chains, IOC indicators, etc.

\bibliography{aaai2026}

\appendix

\section{Dataset construction}
To the best of our knowledge, there is no heterogeneous source dataset based on a large amount of raw CTI that can assist researchers in training appropriate models to improve the efficiency of CTI credibility assessment. 
  To address this problem, we created a new dataset named CTI-200.

We collected 1,000 CTI s from different open-source platforms. 
  We grouped every five CTIs describing the same APT organization into one group, and used one of them as a report to be verified and the other four as clue reports. 
There are a total of 200 groups in our dataset. To ensure the reliability of the dataset, our research team conducted a thorough analysis of each CTI , with the assistance of experts and LLM to ensure that all key entities described in the CTI  and the relationships between related entities were accurately identified and extracted. 
  Considering the existence of redundant and semantically irrelevant information in CTI, we use LLM to process the statements and contents of the report to be verified and obtain its summary. 
For cued CTI, we use LLM to generate concise summaries while simplifying the original reports into cued and interpretable sentences. 
  In order to obtain these interpretable sentences with strong clues, we refer to Cofced \cite{yang2022coarse} based on the ROUGE score \cite{xu2022sequence} to measure the text similarity of clue sentences, and measure the semantic relevance of clue sentences through cosine measurement. 
Those with relevance scores greater than 0.8 are retained and those with relevance scores less than 0.8 are deleted. 
  For negative samples, we use LLM to generate and perform manual detection and modification. 
To increase the difficulty, we follow the principle of maintaining high semantic relevance with positive samples when generating negative samples and delete generated data with low relevance. 
  
Subsequent users can partition the training and testing sets based on specific task requirements using CTI-200 as the foundation. For clarity, we outline our partitioning method as follows: we evenly split CTI-200 into two parts, one for the training set and the other for the validation set. Since CTI-200 has a comprehensive data distribution, the partitioning process can be entirely random. If constructing a knowledge graph is necessary, please build it on the training set to avoid concerns about model reliability due to data leakage.

Finally, it is worth noting that our dataset is constructed in the STIX 2.1 JSON format, which mainly contains the description of the attack event and the labeled organization and time-related descriptions.

\section{Dataset analysis}
\label{appendix:Dataset analysis}
In addition, we provide a detailed data analysis of CTI-200. As shown in Table 10, CTI-200 comprises a total of 1,000 CTI entries, of which 547 are correct and 453 are erroneous. Among these, 200 CTI entries are pending validation, categorized into 117 correct and 83 erroneous entries. Utilizing LLMs, we extracted viewpoints to be tested from these 200 pending CTI entries (for specific details, refer to Section 3.3 of the manuscript). A total of 864 viewpoints were extracted, of which 653 are correct and 211 are erroneous. These viewpoints encompass features related to attack sources (APT), attack behaviors, and temporal characteristics. Additionally, our dataset of 1,000 CTI entries involves 48 APT organizations, of which 44 remain active.

\begin{table}[ht]
\centering
\caption{CTI-200 Data Analysis}
\begin{tabular}{|l|c|c|c|}
\hline
\textbf{Data Analysis} & \textbf{Total} & \textbf{Real} & \textbf{Fake} \\
\hline
CTI & 1000 & 547 & 453 \\
Pending CTI & 200 & 117 & 83 \\
Pending Point & 864 & 653 & 211 \\
APT Related Point & 207 & 177 & 30 \\
Attack Behavior Related Point & 428 & 312 & 116 \\
Timeliness Related Point & 229 & 164 & 65 \\
APT Organization & 48 & - & - \\
Active Organization & 44 & - & - \\
\hline
\end{tabular}
\end{table}

\section{Related work continued}
We believe that mainstream work related to factual reasoning can be divided into two types: one is fact checking and the other is factual QA. \textbf{Factual QA} is usually based on LLMs (such as ChatGPT) for reasoning. Some research focuses on alleviating the hallucination of LLMs\cite{huang2024opera,gunjal2024detecting,leng2024mitigating}, KGs usually contain a lot of reliable information and are therefore considered a solution to alleviate the hallucination of LLMs. Some studies inject KGs with entities as nodes during the training\cite{zhang2022dkplm} or reasoning\cite{jiang2023structgpt} phase, however, they seem to ignore the contextual reasoning capabilities of LLMs, this brings huge cost to the graph traversal of LLMs. Our KG with paragraphs as nodes effectively calls upon the contextual reasoning capability of LLM and reduces the cost of graph traversal.

\section{Experiments continued}
In addition, to evaluate the generalization ability of KGV, we conduct multiple evaluations on the factual QA datasets HotpotQA \cite{yang2018hotpotqa} and 2WikiMQA \cite{ho2020constructing}.

\textbf{HotpotQA} is a dataset based on Wikipedia, containing news published on various platforms and categorizing this news into true and false categories. It includes 113K questions that need verification. There is no need for further processing of these questions. For factual information, we created a factual KG using the same rules.

\textbf{2WikiMQA} is a multi-hop QA dataset that contains structured and unstructured data, as well as evidence information of multi-hop question reasoning paths, where we construct a factual KG using evidence information paragraphs as the structure.

 As shown in Table \ref{tab:t2}, compared with the factual QA method based on LLMs, our KGV improves F1 by at least 2.7\%. Compared with the IRCoT method, our KGV demonstrates more significant advantages. This indicates that our KG-based KGV framework excels in factual reasoning and verification, because the thought chain-based method relies on manually constructed effective example samples, it is difficult and error-prone for specific tasks, and cannot ensure a perfect ``end of reasoning'' in the process of chain reasoning. This may still introduce noisy knowledge and cannot effectively alleviate the LLMs hallucination. As shown in Figure \ref{fig:em-time-density-count}, compared with the KG-based LLM solutions (QKR, KGR, and KGP-T5), our KGV framework outperforms these models with better performance and faster response time due to its integration of a knowledge graph with a simpler structure.

 As shown in Table \ref{tab:t1} and \ref{tab:t2}, our KGV framework also shows strong scalability in fake news detection and factual QA tasks. Although our KGV framework is mainly designed for CTI credibility assessment, it integrates KG and LLMs and also performs well in other factual reasoning tasks.

\begin{table}[h]
    \centering
    \caption{Performance comparison on HotpotQA and 2WikiMQA}
    \label{tab:t2}
    \begin{tabular}{>{\centering\arraybackslash}m{3cm}cc cc}
        \toprule
        \multirow{3}{*}{\centering\textbf{Method}} & \multicolumn{2}{c}{\textbf{HotpotQA}} &\multicolumn{2}{c}{\textbf{2WikiMQA}}\\
        \cmidrule(lr){2-3} \cmidrule(lr){4-5}
        & \textbf{EM} & \textbf{F1} & \textbf{EM} & \textbf{F1} \\
        \midrule
        CoT          & 24.5 & 34.3 & 22.7 & 33.6\\
        IRCoT        & 45.3 & 64.1 & 37.4 & 50.2\\
        DPR          & 43.6 & 62.1 & 35.6 & 51.1\\
        \midrule
        QKR          & 28.0 & 38.1 & 26.4 & 37.5\\
        KGR          & 32.7 & 39.2 & 30.8 & 37.9\\
        KGP-T5       & 46.5 & 66.8 & 39.8 & 53.5\\
        \midrule
        KGV (OURS)  & \textbf{49.2} & \textbf{69.5} & \textbf{43.6} & \textbf{56.5}\\
        \bottomrule
    \end{tabular}
\end{table}

\section{Algorithm}

Algorithm \ref{alg:kgv} describes the process of KGV. $s$ is the verification score, which is used to record the best verification score of the algorithm iteration process; $sim$ is the similarity score of the fragment, which is the semantic similarity used to build the KG currently and is one of the model hyperparameters; $l$ is the fragment extracted by llm in $r$, which is the input of entity extraction; $e$ is the entity extracted by llm in $l$, which is used to retrieve related nodes from $G$;

\begin{algorithm}[h]
\caption{KGV: Knowledge Graph-based Verifier for CTI Credibility Assessment}
\label{alg:kgv}
\begin{algorithmic}[1]
\REQUIRE A CTI report $r$ to be verified, a constructed Knowledge Graph (KG) $G = \{\mathcal{V}, \mathcal{E}\}$ from the CTI-200 dataset, pre-trained LLMs, a set of verification clues $C$, threshold $\theta$ for semantic similarity.
\ENSURE Credibility assessment of the CTI report $r$.
\STATE \textbf{KG Building:} Construct KG using paragraphs as nodes $\mathcal{V}$ and connect nodes $\mathcal{E}$ based on semantic similarity between paragraphs and common keywords.
\STATE Initialize a verification score $s \gets 0$.
\STATE Initialize similarity score $sim$ for segments.
\WHILE{similarity $sim$ < threshold $\theta$  }
    \STATE Extract a segment $l$ from the report $r$ based on relevance.
    \STATE \textbf{Entity Extraction:} Extract entities from the segment using LLMs.
    \FOR{each extracted entity $e$}
        \STATE Retrieve related nodes from $G$ using $e$.
        \IF{related nodes contain evidence}
            \STATE \textbf{Claim Extraction:} Generate claims related to $e$.
            \STATE \textbf{Triple Extraction:} Match claims with triples in $G$.
            \IF{claims match knowledge triples in $G$}
                \STATE Increment verification score $s \gets s + 1$.
            \ENDIF
        \ENDIF
    \ENDFOR
    \STATE $sim \gets sim + 1$
    \IF{$sim$ exceeds threshold $\theta$}
        \STATE Terminate with a positive credibility assessment.
    \ENDIF
\ENDWHILE
\STATE \textbf{Point Verification:} Aggregate the verification results of all claims to determine the overall credibility of $r^{\prime}$.
\RETURN Overall credibility assessment for report $r^{\prime}$.
\end{algorithmic}
\end{algorithm}

\section{Example}
\begin{figure*}[ht]
\centering
 \includegraphics[width=0.8\textwidth]{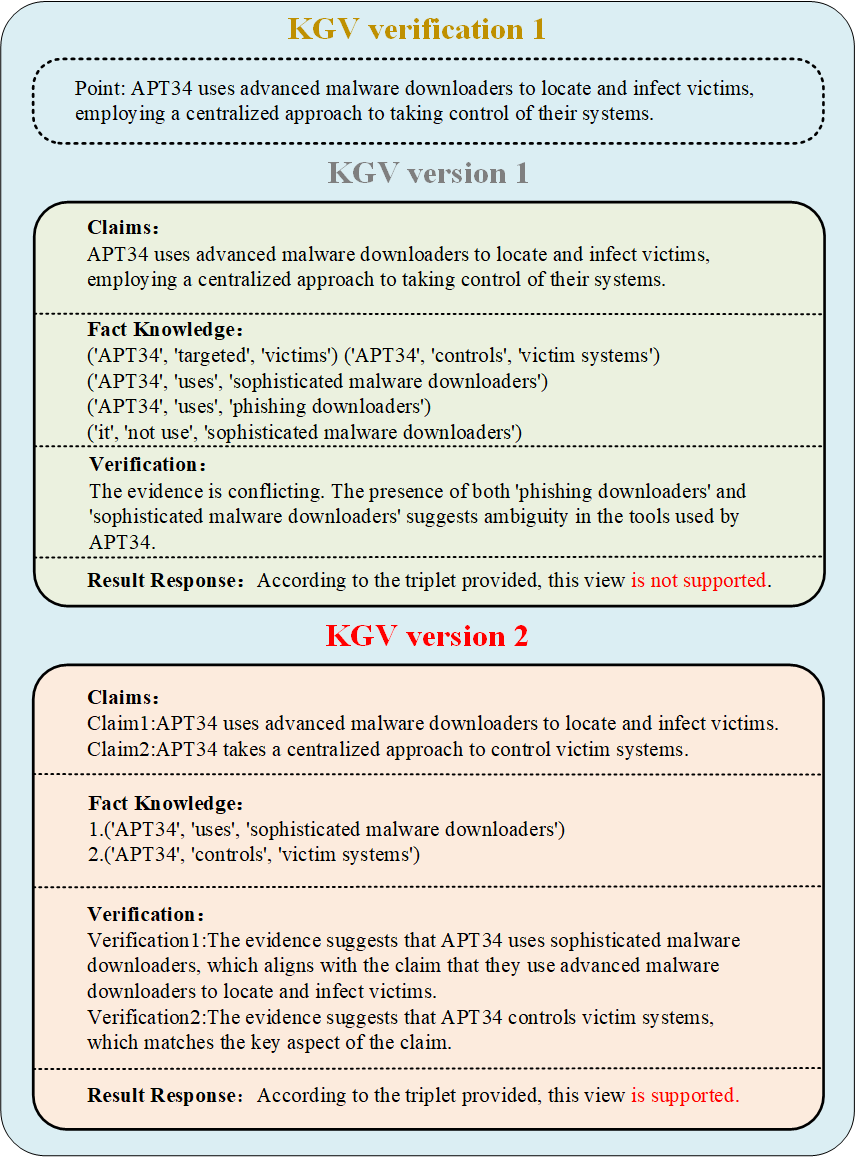}
  \caption{Analysis of KGV verification 1 }
  \label{fig:Analysis_1}
\end{figure*}

We can observe that decomposing opinions into factual statements for fact checking and retrieving factual knowledge through statement triples reduces the number of error cases. 
In the example in Figure \ref{fig:Analysis_1}, in point `` APT34 uses advanced malware downloaders to locate and infect victims, employing a centralized approach to taking control of their systems.'', two different claims to be verified are included:`` APT34 uses advanced malware downloaders locate and infect victims.'' and `` APT34 takes centralized approach to control victim systems.''. In KGV version 1, we followed the original design process, retrieving knowledge through entities in point and verifying directly through the entire point of view. 
LLM is prone to factual illusions and misjudgments. In KGV version 2, we follow the improved design process to fully decompose the point into multiple claims and retrieve factual knowledge by declaring triples. LLM will verify each independent statement to avoid misjudgment.

\begin{figure*}[ht]
\centering
 \includegraphics[width=0.8\textwidth]{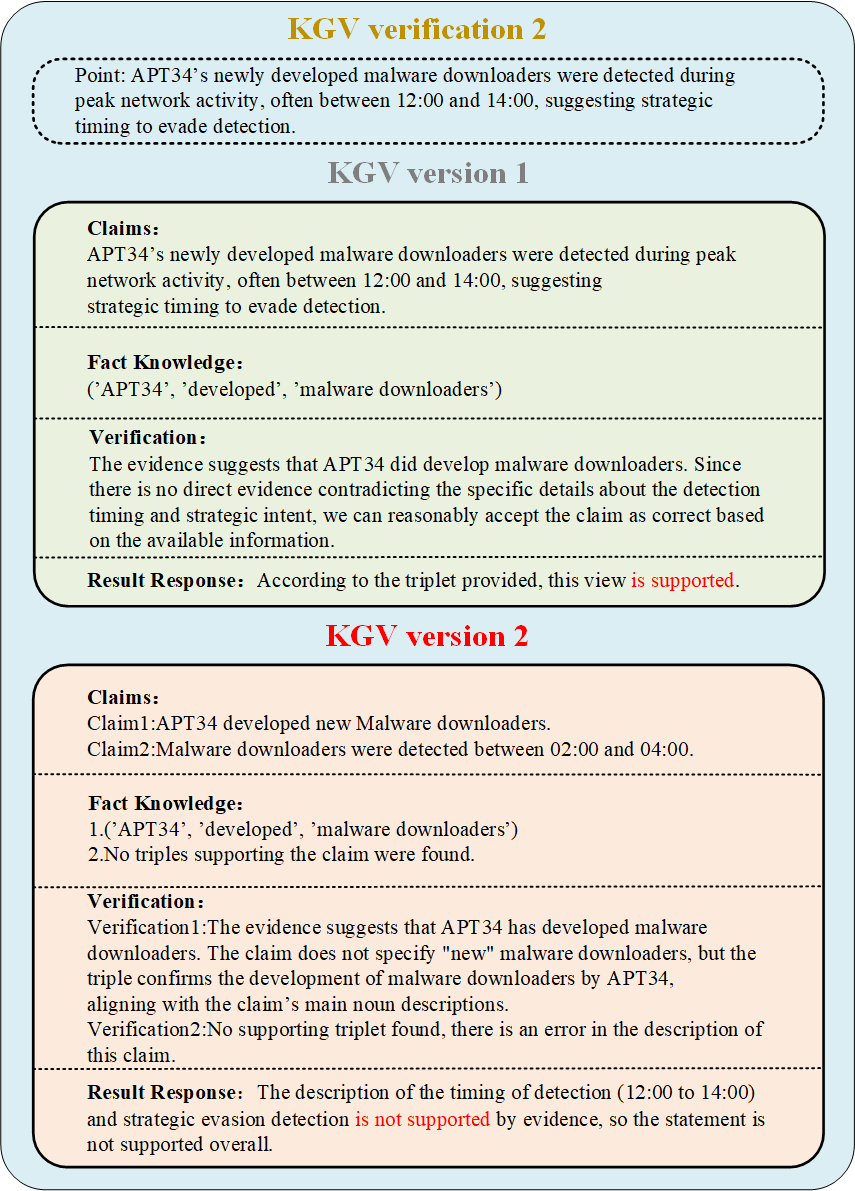}
  \caption{Analysis of KGV verification 2 }
  \label{fig:Analysis_2}
\end{figure*}

The example in Figure \ref{fig:Analysis_2} describes that when the entire opinion is directly verified with a small amount of factual knowledge, the large model will also produce factual illusions and make misjudgments, which also proves the necessity of breaking down opinions into factual statements for fact verification.
  
\begin{figure*}[ht]
\centering
 \includegraphics[width=0.7\textwidth]{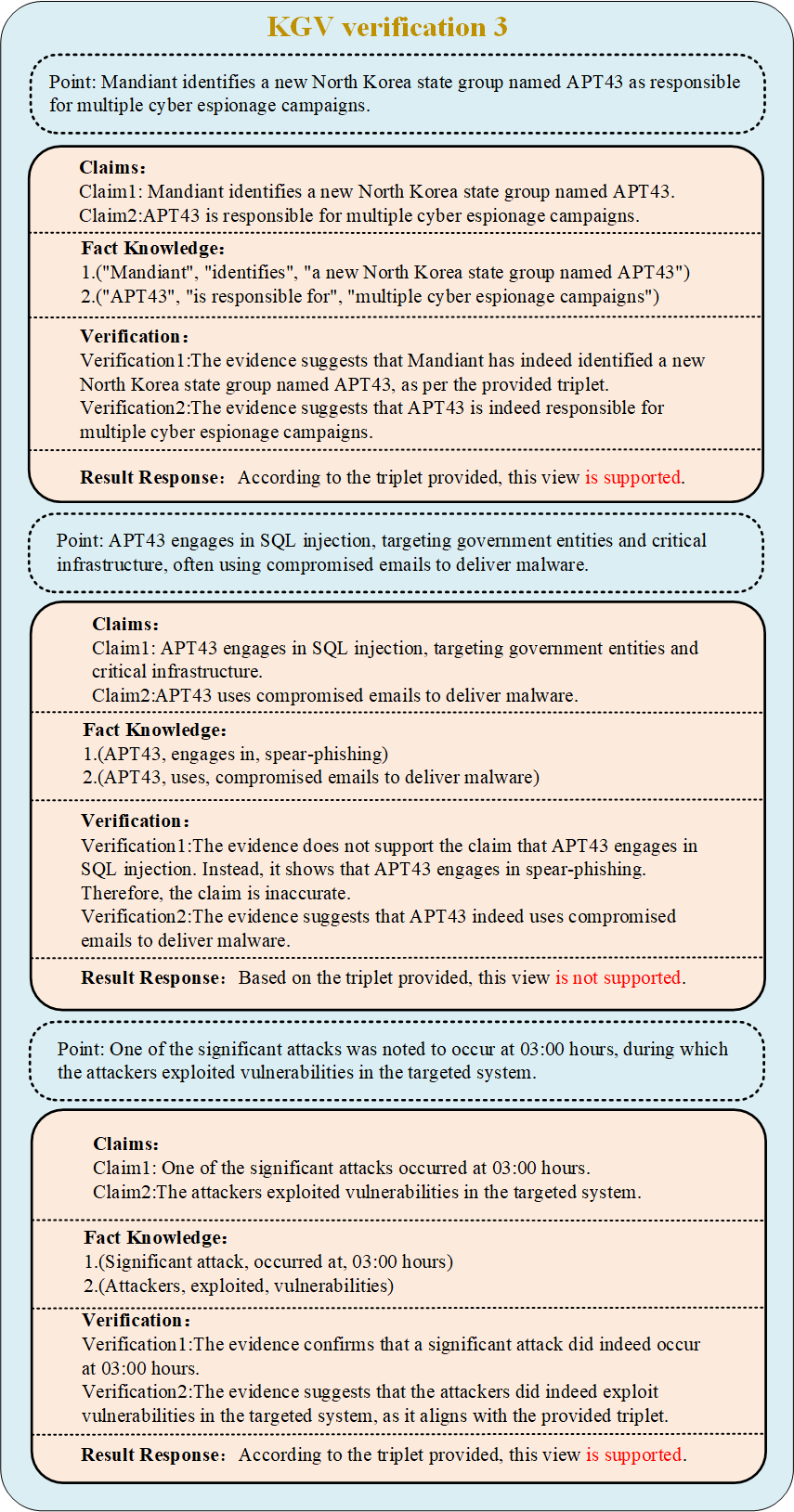}
  \caption{Analysis of KGV verification 3 }
  \label{fig:Analysis_3}
\end{figure*}

\begin{figure*}[ht]
\centering
 \includegraphics[width=0.7\textwidth]{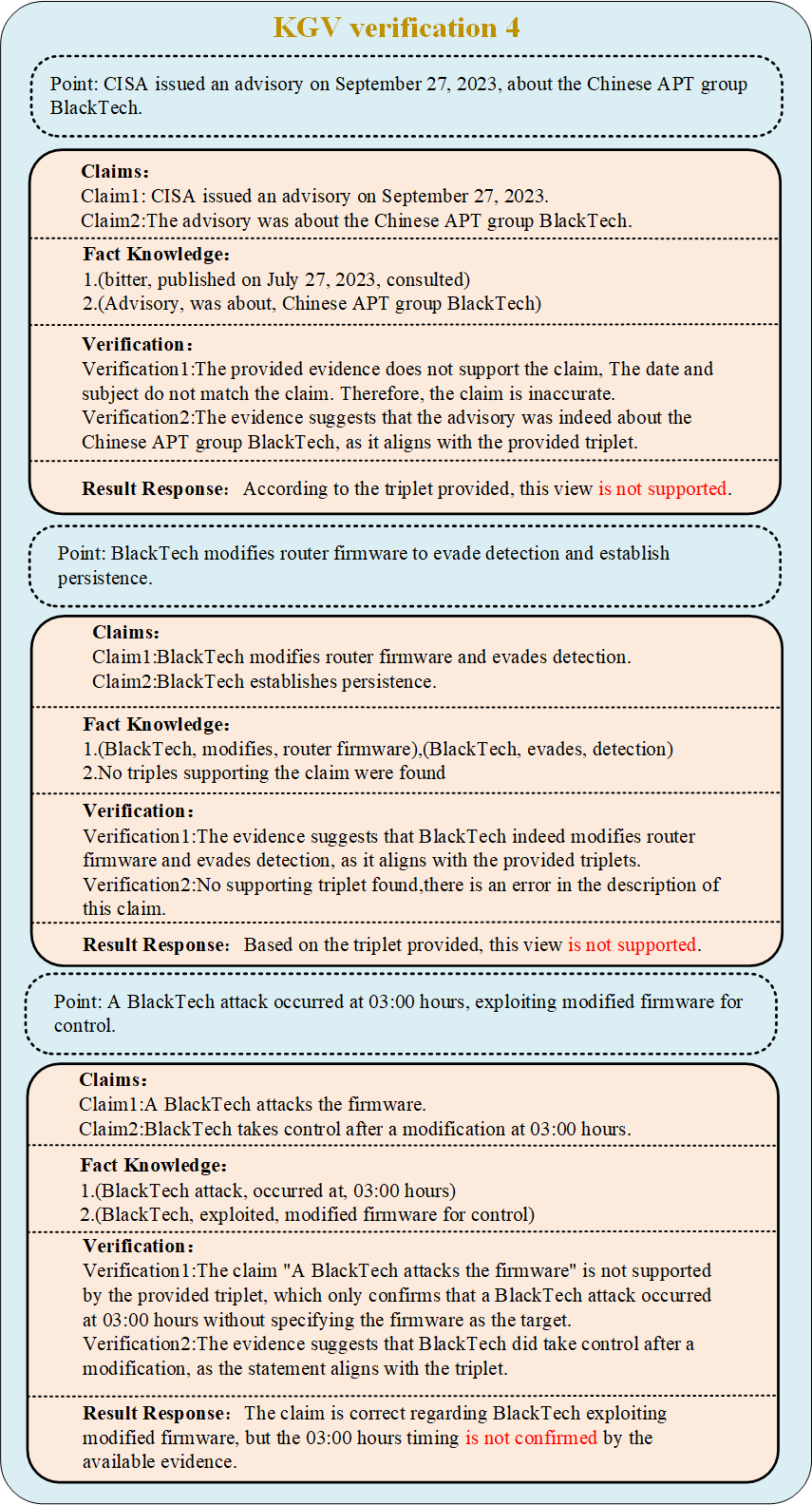}
  \caption{Analysis of KGV verification 4 }
  \label{fig:Analysis_4}
\end{figure*}

\end{document}